\documentclass[preprint,showpacs,amsmath,amssymb,aps,prd,nofootinbib]{revtex4}
\usepackage{epsfig}
\begin{document}

\draft

\title{Extended Uncertainty Relation and \\
Rough Estimate of Cosmological Constant}

\textwidth 16.5cm

\thispagestyle{empty}
\author{Choong Sun Kim\footnote{ E-mail : cskim@yonsei.ac.kr}}
\affiliation{ Space-time and Symmetry Convergence Center (SSCC), Yonsei University, Seoul 120-749, Korea \\
}

\begin{abstract}

\noindent One brief idea on the extended uncertainty relation and the  dynamical quantization of space-time at the Planck scale is presented.
The extended uncertainty relation could be a guiding principle toward the renormalizable quantum gravity.
Cosmological constant in the Universe as a quantum effect is also roughly estimated.
\\


\end{abstract}
 \maketitle \thispagestyle{empty}
%
\textwidth 18cm

In the classical theory of general relativity \cite{GR} the distribution and
motion of matter ($i.e.$
energy-momentum) are determined by gravitational field functions ($i.e.$
description of space-time), and at the same time gravitational field functions
are determined by the distribution and motion of matter through Einstein's
field equation.  In the modern quantum theory space-time is closely related
to energy-momentum of matter by Heisenberg's uncertainty
relation \cite{UP}.  I think it is unreasonable to treat
space-time and matter separately and to quantize just the metric tensor and
not space-time itself in  quantum gravity theory \cite{QG}.
{\it E.~g.},   in the extended
structure of a fiber bundle,
where the space-time manifold has  not only a charged space at each
point but also a tangent space, they just concern the quantization of the
charged space but the tangent space.

In this brief comment, first I would like to extend the Heisenberg's uncertainty principle to a relation between
space-time and energy-momentum tensors, henceforth to apply directly to Einstein's field equations to quantize
the classical gravity. As in the process of development of classical quantum mechanics in early 20 century,
the procedure would be kind of ad hoc and brutish but, I hope, contain some hidden truth to find further
more refined theory of quantum gravity as well as to guide us to the deep insight of space-time at the Planck scale.
Then, by matching the classical value of energy-momentum tensor to the quantum expectation value
of the operator, and by estimating the first order quantum effects, I would like to calculate
the cosmological constant (dark energy) of our Universe.

Now I propose  an extended uncertainty relation, and study its implication to Einstein's
general relativity.
\begin{itemize}
\item{1.} Extend the Heisenberg's uncertainty principle,
\begin{equation}
[\hat X^{\mu\nu}(x),\,\hat T_{\alpha\beta}(x^\prime)] =
i\hslash\,\delta^{\mu\nu}_{\alpha\beta}\delta^{(4)}(x-x^\prime) \>,
\end{equation}
where $\hat X^{\mu\nu}$ is the space-time tensor operator, and $\hat T_{\alpha\beta}$ is
the energy-momentum tensor operator which is connected to the energy-momentum tensor in
Einstein's classical general relativity equation,
\begin{equation}
G_{\mu\nu}\equiv R_{\mu\nu} - 1/2 g_{\mu\nu}R = {8\pi G\over c^4} T_{\mu\nu}.
\end{equation}
The relation between the first row and the first column of  Eq. (1) is the same as the original
Heisenberg's uncertainty principle,
\begin{eqnarray}
[\hat x^\mu(x),\ \hat p_\nu(x^\prime)] &= i\hslash\,\delta^\mu_\nu \delta^{(4)}(x-x^\prime).
\end{eqnarray}
\item{2.} Now $\hat T_{\alpha\beta}$, as an operator, from the generalized uncertainty principle Eq. (1),
will be defined as
\begin{eqnarray}
\hat T^{\mu\nu} = -i\hslash{\partial\over\partial \hat X_{\mu\nu}}  ~~~~
\Rightarrow ~~~~ \hat T^{\mu\nu} = -i\hslash\nabla_{X_{\mu\nu}}\>,
\end{eqnarray}
with covariantization (for general invariance),
which is meaningful after applied to the wave-function of the space-time at the Planck scale,
\begin{equation}
\hat T^{\mu\nu} \Phi_{\rm Univ} = A^{\mu\nu} \Phi_{\rm Univ},
\end{equation}
where $\Phi_{\rm Univ}$ is a wave function for the quantized space-time, and $A^{\mu\nu}$ is the corresponding eigenvalue of energy-momentum.
In practise for the present low energy scale, we may approximate a smooth whole universe wave-function,
or a symmetric wave-function within Schwarzschild radius.
\item{3.} And the classical value  $T_{\mu\nu}$ of Eq. (2) can be calculable as an averaged expectation value of
the operator $\hat T_{\mu\nu}$ as,
\begin{equation}
T_{\mu\nu}^{\rm CL}=\langle \hat T_{\mu\nu} \rangle = \langle \Phi_{\rm Univ} | \hat T_{\mu\nu}  | \Phi_{\rm Univ} \rangle
 = T_{\mu\nu}^{(0)} + T_{\mu\nu}^{(1)} + ...
\end{equation}
Here $T_{\mu\nu}^{(0)}$ would be the first order approximation of $T_{\mu\nu}^{\rm CL}$
without any quantum correction. We presumed the validity of perturbation in (6) by assuming a smooth (no singularity)
macroscopic wave-function of the Universe with weak gravity.
\item{4.} Through the quantum effects,  Eq. (2) is modified to
\begin{equation}
G_{\mu\nu} = {8\pi G\over c^4} [T_{\mu\nu} + T_{\mu\nu}^{\rm QE}] \approx
{8\pi G\over c^4} [T_{\mu\nu} + T_{\mu\nu}^{(1)}],
\end{equation}
which modifies the structure of space-time from the classical gravity.
\item{5.} Here $T_{\mu\nu}^{(1)}$ is the first order quantum effect,
which corresponds to energy-momentum tensor of the vacuum as
\begin{equation}
T_{\mu\nu}^{(1)} \simeq \rho_{_{\rm VAC}} g_{\mu\nu},
\end{equation}
calculable from the quantum correction of the vacuum polarization
of space-time.
From Eq. (7), $T_{\mu\nu}^{(1)}$ can also be interpreted as the cosmological constant $\Lambda$ from
$$ G_{\mu\nu} + \Lambda g_{\mu\nu} = {8\pi G\over c^4} T_{\mu\nu},$$
where $$\Lambda \simeq - {8\pi G\over c^4} \rho_{_{\rm VAC}}.$$
\item{6.} From Eq. (1) the operator $\hat X^{\mu\nu}$ can be defined as
\begin{eqnarray}
\hat X^{\mu\nu} = i\hslash{\partial\over\partial \hat T_{\mu\nu}} ~~~~
\Rightarrow ~~~~ \hat X^{\mu\nu} = i\hslash\nabla_{T_{\mu\nu}}\>,
\end{eqnarray}
which can give the quantization of space-time itself near around strong gravitational fields, $e.g.$
within the horizon of a black-hole, just like the energy quantization of a quantum mechanical bounded state.
This procedure might be related to ``loop quantum gravity" \cite{LQG}, which proposes
Planck scale granularity of space-time  based on the background independence and the diffeomorphism
invariance.
The difference from the loop quantum gravity would be that in our approach space-time is continuous in free space,
and is quantized dynamically with the gravitational interaction.
We note that this dynamic space-time quantization could be experimentally tested possibly through
gravitational lenz effects of a black-hole, or through early universe astroparticle observables. It may also affects the black-hole thermodynamics.
\end{itemize}

Now let us estimate numerical size of (possibly dark energy)
$\rho_{_{\rm VAC}}$ (and the size of the cosmological constant).
\begin{itemize}
\item{1.} As a rough numerical approximation, I guess that
taking only the vacuum polarization of graviton in weak Abelian gravitational field
would be rather a good approximation just for the cosmological constant, which is  macroscopic
galactic feature of the Universe.
\item{2.} Following Lamb shift of Hydrogen spectrum \cite{Lamb}
$$
|\delta E_{_{\rm Lamb}}| \simeq \alpha^5 {m_e c^2 \over 30 \pi},
$$
the quantum correction of the vacuum energy density
of the Universe is approximately,  considering Planck mass fluctuation,
\begin{eqnarray}
|\delta \rho_{_{\rm VAC}} | &\sim&  \left({M_*^2\over M_P^2}\right)^5 \frac{M_P c^2}{30 \pi}{1 \over L_P^3} \\
&\simeq& \left({M_*^2\over M_P^2}\right)^5 0.4\times 10^{92}~g/cm^3~,
\end{eqnarray}
where $M_P$ and $L_P$ are Planck mass and Planck length, respectively.
Here $M_*$ is the scale of observation, therefore,
\begin{eqnarray}
|\delta \rho_{_{\rm VAC}} | &\sim& 10^{-98}~g/cm^3~~~~~~~~~{\rm for}~~M_*={\cal{O}}(1)~{\rm GeV}, \nonumber\\
&\sim& 10^{-153}~g/cm^3~~~~~~~~{\rm for}~~M_*={\cal{O}}(1)~{\rm keV}, \nonumber
\end{eqnarray}
which are within the experimental upper bound \cite{CC-exp}
$$|\delta \rho_{_{\rm VAC}}^{\rm exp} | < 4 \times 10^{-29}~g/cm^3,$$
where the upper limit corresponds to $M_* \sim 10^{7}$ GeV.
\item{3.} And finally the cosmological constant $\Lambda$ is
\begin{eqnarray}
\Lambda =  {8\pi G\over c^4} \rho_{_{\rm VAC}}. \nonumber
\end{eqnarray}
\end{itemize}



Since space-time is closely related to matter through Einstein's
field equation and Heisenberg's uncertainty relation,
I proposed an extended uncertainty principle to the relation between space-time and
energy-momentum tensors, which could not only describe the Universe wave function
but also quantize space-time itself through the relation.
This extended uncertainty relation could be a guiding principle toward the renormalizable quantum gravity,
or toward the resolution of landscape problem.
Because at the energy scale of order $10^{15-16}$~GeV (or equivalently
distance scale of order $10^{-29}$~cm) the gravitational quantum effects appear just
around the corner, at the present experimental level we have no way to
investigate directly quantum gravity.  However, I proposed even
at the scale of GeV, though space-time has little curvature
($i.e.$ by Einstein's field equation,
$g_{\mu\nu} = \delta_{\mu\nu} + h_{\mu\nu},\
h_{\mu\nu}$ small), space-time itself is quantized in a
measurable sense through the quantum effects of the cosmological constant.
And its dynamical space-time quantization could be experimentally tested
through gravitational lenz effects of a black-hole as well.
Please also note that recently BICEP2 experiment \cite{BICEP2} has detected primordial gravitational waves, which may imply the generation of large scale curvature perturbation.
\\

\noindent {\bf Acknowledgement:}

\noindent Discussions with J. H. Cho, P. Frampton, S. J. Hyun are gratefully acknowledged.
The work was supported by the National Research Foundation of Korea (NRF)
grant funded by Korea government of the Ministry of Education, Science and
Technology (MEST) (No. 2011-0017430) and (No. 2011-0020333).

\end{document}